\documentclass[aps,prb,twocolumn,groupedaddress,showpacs]{revtex4}
\usepackage{graphics} 
\usepackage{epsfig}
\usepackage{graphicx}
\usepackage{dcolumn}

\begin{document}

\title{Quasi-ballistic, nonequilibrium electron distribution in
  inhomogeneous semiconductor structures}

\author{Dan Csontos}
\email{csontos@phy.ohiou.edu}
\affiliation{Department of Physics and Astronomy, and Nanoscale and
  Quantum Phenomena Institute, Ohio University, Athens, Ohio 45701, USA} 
\author{Sergio E. Ulloa}
\affiliation{Department of Physics and Astronomy, and Nanoscale and
  Quantum Phenomena Institute, Ohio University, Athens, Ohio 45701, USA} 

\begin{abstract}
{\footnotesize We report on a study of quasi-ballistic transport in
  deep submicron, inhomogeneous semiconductor structures, focusing on
  the analysis of signatures found in the full
  nonequilibrium electron distribution. We perform self-consistent 
numerical calculations of the
  Poisson-Boltzmann equations for a model $n^{+}-n^{-}-n^{+}$ GaAs
  structure and realistic, energy-dependent scattering. We show that, 
in general, the electron distribution displays significant, temperature 
dependent broadening  
  and pronounced structure in the high-velocity tail of the
  distribution. The observed
  characteristics have a strong spatial dependence, related
  to the energy-dependence of the 
scattering, and the large inhomogeneous electric field variations in 
these systems. We show that in this
  quasi-ballistic regime, the high-velocity tail structure is due to 
pure ballistic transport, whereas 
the strong broadening is due to electron scattering within the
  channel, and at the source(drain) interfaces.}  
\end{abstract}

\maketitle

As the downscaling of semiconductor devices has reached well into the
deep submicrometer regime, the nonequilibrium nature of the carrier
dynamics has become exceedingly important. In these systems, channel
lengths comparable to the electron mean free path are possible at room
temperature, and thus, ballistic and hot electron effects are expected
to strongly influence the transport properties of modern
devices. Although numerous studies of hot electron and ballistic 
transport have been performed,\cite{ravaioliSST98} 
there have been very few works
on the microscopic carrier dynamics. It is, e.g., not clear how the
details of the scattering at the channel interfaces, as well as within
the channel, affect the ballistic nature of the transport, what
the microscopic signatures of quasi-ballistic transport are, and how
these issues are responsible for limiting the current in these
systems.\cite{lundstromIEEE02, svizhenkoIEEE03} 

A comprehensive microscopic analysis of these issues requires the study of the
electron distribution function, which is a challenging task, both
experimentally  
and theoretically, due to the strongly nonequilibrium nature of the problem.
There have been a few recent theoretical studies of ballistic MOSFETs 
\cite{lundstromIEEE02, rhewSSE02} in
which the electron distribution was obtained by solution of the
Boltzmann Transport Equation (BTE) in the collisionless limit. In this
pure ballistic limit, according to the models, the electron distribution
within the channel is composed of near-equilibrium electron distributions 
in the contacts,
and thus, due  to the inhomogeneous self-consistent field, only displays
an asymmetric, pronounced high-velocity peak, corresponding to ballistic
electrons. In recent experimental studies on the other hand,\cite{experiments}
the
electron distribution in submicron, inhomogeneous III-V systems was
measured using Raman spectroscopy, and was shown to display strongly
broadened velocity distributions and interesting high-velocity tail
structure. It is clear that in realistic room-temperature devices, the
electron scattering rate is finite and thus, a full solution of the BTE,
including scattering is required. Such an analysis was in fact already
initiated by Baranger and Wilkins.\cite{baranger}
These authors solved the BTE self-consistently with the Poisson equation
for $n^{+}-n-n^{+}$ GaAs structures, and found significantly
out-of-equilibrium electron distributions in the channel region, and
in particular, a high-velocity peak corresponding to ballistic
electrons. However, the physical mechanisms of the strong broadening
and the details of the electron distribution characteristics were not
clearly resolved. In addition, the scattering introduced in the BTE
was structureless, while the distribution function in inhomogeneous
systems is strongly energy-dependent. 
~\\
~\\
In this paper we present a numerical analysis of the nonequilibrium
electron distribution in submicron, inhomogeneous semiconductor
structures. We show that the electron distribution function has a
highly nonequilibrium form, is significantly broadened, and displays
interesting structure in the high-velocity tail. In addition, the
observed features are spatially dependent and very sensitive to the
inhomogeneous electric field, temperature and detailed characteristics
of the scattering, where we find that an energy-dependent scattering
process strongly affects the details of the electron distribution. In
addition to providing insight to the general
quasi-ballistic transport 
characteristics in the electron distribution of submicron structures,
we believe that our results are relevant for the recently observed
nonequilibrium velocity distributions observed in the recent
experiments. 

Our study is based on numerical analyses in
which we explictly calculate the 1D, steady-state electron
velocity distribution function, $f(x,v)$, through a model
$n^{+}-n^{-}-n^{+}$ GaAs structure.\cite{parameters} We consider a nondegenerate system and 
use a numerical, self-consistent approach
based on the coupled Poisson-Boltzmann equations, which enables us to
capture essential nonequilibrium and inhomogeneous transport
phenomena.\cite{csontosJCE04} Scattering is treated within the
relaxation-time approximation, using realistic, energy-dependent scattering
rates corresponding to polar-optical phonon (POP) scattering, which is the
dominating scattering mechanism in GaAs at room temperature.\cite{scattering}
The POP scattering rate is given by\cite{nag}

\begin{eqnarray}
\frac{1}{\tau_{p}(\varepsilon)} & = & \frac{e^{2}\omega_{0}\kappa}{2^{3/2}\pi
  \epsilon_{0} \hbar} 
  \sqrt{\frac{m^{\ast}}{\varepsilon}}  \{ n_{0}\sinh^{-1}\left
  [\frac{\varepsilon   }{\hbar \omega_{0}}\right ]^{1/2} \nonumber \\
 & + & (n_{0}+1)\sinh^{-1}\left [\frac{\varepsilon}{\hbar \omega_{0}}-1
  \right ] ^{1/2}  \}~, 
\label{pop}
\end{eqnarray}
where $\kappa=(1/ \epsilon_{\infty} -1/\epsilon)$, $\epsilon_{\infty}$
is the high-frequency dielectric constant, $\epsilon$ is the static
dielectric constant, $\hbar\omega_{0}$ is the optical phonon energy 
and $n_{0}=[\exp(\hbar\omega_{0}/k_{B}T)-1]^{-1}$ is the phonon
occupation number. In Eq. (\ref{pop}) the first term corresponds to
POP absorption and the second to POP emission, which only occurs when
$\varepsilon(p)>\hbar \omega_{0}$. 

\begin{center}
\begin{figure}[h]
\scalebox{0.45}{\epsfig{file=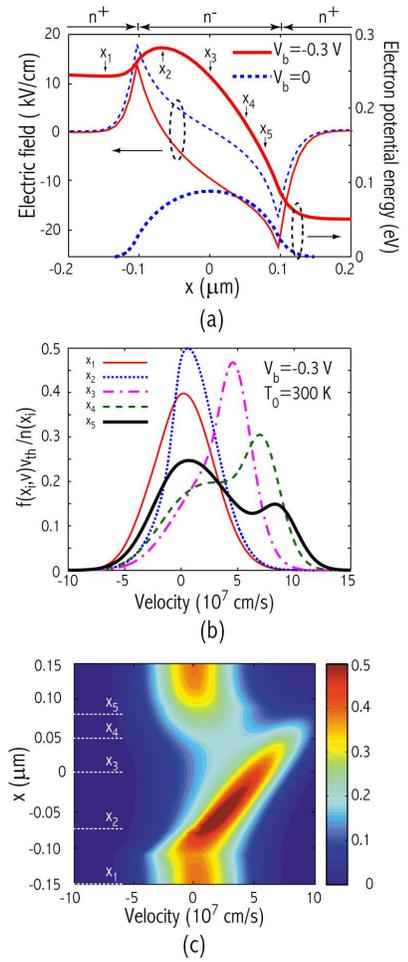}}
\caption{\label{figure1} (color online) (a) Electric field (thin
  lines) and potential 
  energy (thick lines) profile for $V_{b}=0$ (dashed lines) and
  $V_{b}=-0.3$ V (solid lines), (b) normalized electron distribution
  calculated at the 
  spatial points denoted by the arrows in (a), and (c) contour plot of
  $f(x,v)v_{th}/n(x)$ around the $n^{-}$ region for a
  $n^{+}-n^{-}-n^{+}$ GaAs structure with dimensions 1.5/0.2/1.5
  $\mu$m and doping profile $10^{17}/10^{13}/10^{17}$
  cm$^{-3}$. Scattering time is $\tau_{2}=2.5\cdot 10^{-13}$ s.}  
\end{figure}
\end{center}

In Fig.\ \ref{figure1} we show the electron potential
energy, electric field and distribution function
at $T=300$ K, and $V_{b}=-0.3$ V, for a 200 nm long lightly doped
($10^{13}$ cm$^{-3}$) GaAs slab sandwiched 
between two highly doped ($10^{17}$ cm$^{-3}$) long
contacts. In order to highlight the effects of inhomogeneities and
scattering while keeping the nature of the scattering {\em structureless},
we first use constant scattering times in our calculations (in
Fig.\ \ref{figure1}, 
$\tau_{2}=2.5\cdot 10^{-13}$ s), and subsequently
compare these results with realistic energy-dependent
scattering times. For comparison, the potential energy and
electric field at $V_{b}=0$ are plotted in
Fig.\ \ref{figure1}(a) (dashed lines). Several direct observations can
be made from 
the electrostatics depicted in Fig.\ \ref{figure1}(a). A 
potential barrier is formed in the $n^{-}$ region due
to the inhomogeneous doping profile that gives rise to the diffusion
of electrons from the highly doped $n^{+}$ into the lightly doped $n^{-}$
region. Second, as a result, a strong electric field on the order of
10 kV/cm is formed within the device, even in the absence of an
external field. Third, as a bias voltage is applied, a large
portion of the voltage drop occurs over the submicron $n^{-}$ region,
giving rise to a strongly inhomogeneous field distribution, in
contrast to the $n^{+}$ contact regions, where the field in comparison
is very low and constant. 

In Fig.\ \ref{figure1}(b), the calculated normalized electron
distribution function, $f(x_{i},v)v_{th}/n(x_{i})$, where $v_{th}$ is
the thermal velocity, is shown at different spatial points along 
the sample for the $V_{b}=-0.3$ V case. At $x_{1}=-0.15$ $\mu$m, the
electrons injected 
from the source display a shifted MB distribution. This is expected
throughout the highly-doped contact regions,  
where the field is constant and low ($\approx$ 0.1 kV/cm) and
thus, the distribution function can be described by the linear
response solution to the homogeneous BTE,
$f(x,v)=f_{MB}(x,v)[1-veE(x)\tau/k_{B}T]$. At 
$x_{2}=-0.07$ $\mu$m, corresponding to the top of the
potential barrier, the distribution is asymmetric, showing a
suppression of electrons in the $v<0$ tail of the distribution. This is
caused by the asymmetric potential barrier, which prevents left-moving
electrons ($v<0$) to reach $x_{2}$. Deep in the $n^{-}$ region, for $x>x_{2}$,
the electron distribution displays a strong spatial dependence as well
as significant broadening. To better understand the overall spatial
dependence of the distribution 
function, Fig.\ \ref{figure1}(c) shows the contour plot of $f(x,v)$,
around the interesting $n^{-}$ region. It is clear
that the electron distribution in the $n^{-}$ region displays a
distinct peak that is rapidly shifted toward higher velocities for
increasing $x>x_{2}$. This is a signature of {\em ballistic transport} and
occurs due to the rapid and large (compared to $k_{B}T$) potential energy drop
\cite{baranger, lundstromIEEE02, rhewSSE02} and corresponding strong
inhomogeneous  
electric field. In addition to the ballistic peak, for
increasing $x>x_{2}$, the electron distribution is broadened and a
low-velocity contribution gradually builds up until the distribution
is dominated by a low-velocity peak, see point $x_{5}$ in
Fig.\ \ref{figure1}(b) and the spatial dependence in
Fig.\ \ref{figure1}(c). The origin of this broadening is two-fold: The
thermionically injected electrons at $x_{2}$, with $v>0$, are 
gradually thermalized at a rate $1/\tau$ as they cross  the channel,
thus gradually ``depopulating'' the high-velocity peak 
 corresponding to
ballistic electrons.
Second, electrons backscattered at the
channel-drain interface, as well as injected electrons with $v<0$ from a
near-equilibrium distribution in the drain,
penetrate the channel with negative velocities, contributing to the
negative and low-velocity end of the electron distribution. These
electrons are scattered at a rate $1/\tau$, but also experience 
scattering at the steep potential energy barrier close to the
source. Hence, close to the 
source and deep in the $n^{-}$ region, the distribution is dominated by
the ballistic peak corresponding to the source-injected ballistic
electrons [see Fig.\ \ref{figure1}(b)]. Close to the drain, however,
where the effective potential barrier is lower, the drain-injected and
reflected electrons dominate the distribution and obscure the
ballistic peak.   

\begin{figure}[h]
\scalebox{0.4}{\epsfig{file=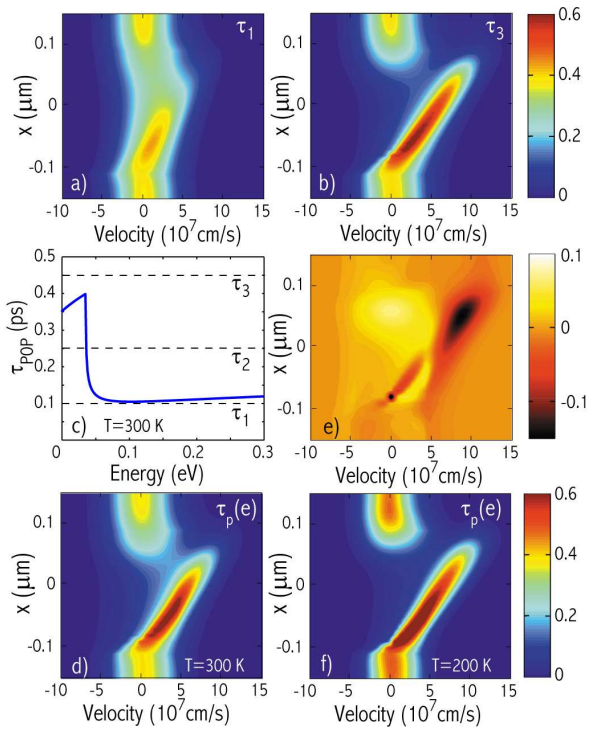}}
\caption{\label{figure2} (color online) Normalized distribution function,
  $f(x,v)v_{th}/n(x)$, for $\tau_{1}=1.0\cdot 10^{-13}$ s (a),
  $\tau_{3}=4.5\cdot 10^{-13}$ s (b), $\tau_{p}(T=300$ K) (d), and
  $\tau_{p}(T=200$ K) (f). (c) POP scattering time at $T=300$ K. (e)
  Difference between the normalized distribution functions calculated
  for $\tau_{p}(T=300$ K) and $\tau_{3}$ [plots (d) and (b)].}
\end{figure}

The spatial dependence of the electron velocity distribution is
strongly influenced by the characteristics of the external and built-in
fields, mobility, scattering mechanisms and temperature. In 
Fig.\ \ref{figure2}(a), (b) we show the
distribution function calculated for the constant scattering times
$\tau_{1}=1.0\cdot 10^{-13}$ s and $\tau_{3}=4.5 \cdot 10^{-13}$ s, for the
same structure and system parameters studied in Fig.\ \ref{figure1}
(with $\tau_{2}=2.5 \cdot 10^{-13}$ s). While energy-independent, 
these scattering times correpond to realistic mobilities of GaAs. For
the short scattering time, $\tau_{1}=1.0\cdot 10^{-13}$ s,
the ballistic peak structure seen in Fig.\ \ref{figure1}(c) is significantly
less pronounced and the average velocity of the electrons is strongly
reduced. The latter observation is a natural consequence of the increase 
of the scattering rate. However, in addition, the fraction of the potential
drop that occurs over the channel region also decreases with an increase in 
the scattering rate (not shown here) and hence, due to the decreased effective 
potential barrier, the drain-injected and reflected electrons 
penetrate deep into the channel. Therefore, the resulting distribution, 
although strongly out-of-equilibrium, does not display distinct ballistic 
features, but becomes broadened, asymmetric and shifted. For
$\tau_{3}=4.5\cdot 10^{-13}$ s  
[Fig.\ \ref{figure2}(b)], the opposite
effects are observed: A very distinct ballistic peak dominates the distribution
function deep into the $n^{-}$ region close to the drain, and less broadening 
is observed. This is due to an increase in the electron velocity
given by the decreased scattering rate, and an increase in the
effective potential 
barrier seen by electrons with $v<0$ coming from the drain.

In realistic systems, the scattering and relaxation processes can have a 
strong energy-dependence and thus, it is interesting to consider the effects
of an energy-dependent scattering rate on the nonequilibrium
distribution function. 
In Fig.\ \ref{figure2}(c) the energy-dependent POP scattering time
$\tau_{p}(\varepsilon)$, 
  calculated from Eq. (\ref{pop}) at $T=300$ K, is shown. For
  clarity, in the figure, we 
  have also indicated the values of the constant scattering times used
  in the calculations shown in Figs.\ \ref{figure2}(a), (b) and
  \ref{figure1}(c). Below the {\em threshold} for POP emission, the
  scattering time is in the range $\approx 3.5-4\cdot 10^{-13}$ s,
  whereas above the threshold, the scattering time has a relatively
  weak dependence on energy (in this particular energy interval) and is 
close to $\approx 1\cdot 10^{-13}$
  s. The corresponding electron distribution function, calculated at
  $T=300$ K is shown in Fig.\ \ref{figure2}(d). Strong signatures
  of quasi-ballistic transport are seen, i.e., formation of a
  ballistic peak that evolves toward higher velocities, similarly to
  the calculations performed with the constant, energy-independent
  scattering times, and significant broadening and low-velocity contributions
due to the scattering processes discussed above. The details of the
distribution  
function and its spatial dependence are, however, different. In 
Fig.\ \ref{figure2}(e), the difference between the normalized
  functions
  $f[x,\tau_{p}(\varepsilon)]v_{th}/n(x)-f(x,\tau_{3})v_{th}/n^{\prime} (x)$  
[$n(x)$ and $n^\prime (x)$ are the electron densities corresponding to
  the two calculated  
distribution functions] is shown. The observed
difference can be explained as follows: Electrons with velocities below 
$v_{tp}\approx 4.3\cdot 10^{7}$ cm/s, which corresponds to a kinetic
energy comparable  
to the threshold for POP emission [Fig.\ \ref{figure2}(c)], have a large mean 
free path and behave
  ballistically, as concluded by the results shown by the calculations
  for $\tau_{3}$ shown in Fig.\ \ref{figure2}(b). For electrons with
  higher kinetic energies on the other hand the scattering rate is
  higher and thus the electrons relax toward lower energies at an
  increasing rate. However, the picture is more complicated, since the detailed
energy-dependence of the scattering effectively also changes the characteristics
of the inhomogeneous field due to the charge redistribution caused by
the scattering.  

Finally, we briefly discuss the effects of
  temperature, which is the parameter that, next to the applied
  electric field, is most easily tuned in experiments. In
  Fig.\ \ref{figure2}(f) we show the
  electron distribution calculated at $T=200$ K. Below the threshold
  for POP emission, the calculated scattering time (not shown here) 
is in the range $\approx
  8.2-9.2\cdot 10^{-13}$ s, whereas above threshold, the scattering
  time has weak energy dependence and is $\approx 1.5\cdot
  10^{-13}$ s. As anticipated
  from the above discussion, the main effects of lowering the
  temperature and, thus, increasing of the phonon scattering time, are:
  $i)$ higher average velocities, $ii)$ larger spatial 
  extension of the ballistic peak, $iii)$ narrowing of the
  ballistic 
  peak in the n$^{-}$ region as well as the diffusive peaks in the
  source/drain, and $iv)$ suppression of the drain-injected electron 
  contribution to the electron distribution in the channel region, due
  to an increase in the effective potential barrier.     

In summary, we have studied the electron distribution in 
inhomogeneous, deep submicron semiconductor structures by self-consistent
calculations of the semiclassical BTE. We have shown that the electron
distribution in general is strongly out-of-equilibrium, significantly broadened 
and displays pronounced structure in the high-velocity tail. These characteristics
of quasi-ballistic transport are very sensitive to the energy-dependent scattering
in the channel and at the source(drain) interface, as well as the strongly inhomogeneous
electric field. Our results are similar to recent experimental
observations. 

{\em Note added after submission:} An interesting study of quasiballistic
transport in nanoscale semiconductor structures with focus on the scattering 
and the mathematical nature of the BTE at the top of the potential energy barrier 
at the source-channel interface has been published very recently 
by Sano.\cite{sanoPRL04, sanoAPL04} We note that 
although the calculations were done for completely different material (Si)
and system parameters, the features of the distribution shown in Fig. 3 
of Ref. \onlinecite{sanoAPL04} are similar to the features of the distributions
calculated here, and are due to the mechanisms discussed in our paper. \\

This work was supported by the Indiana 21st Century Research and
Technology Fund.


\begin{thebibliography}{99}
\bibitem{ravaioliSST98} See, e.g., U. Ravaioli, Semicond. Sci. Technol. 
{\bf 13}, 1 (1998), and references therein.
\bibitem{lundstromIEEE02} M. Lundstrom, and Z. Ren, IEEE Trans. Electron
Devices {\bf 49}, 133 (2002).
\bibitem{svizhenkoIEEE03} A. Svizhenko, and M. P. Anantram, IEEE Trans. 
Electron Devices {\bf 50}, 1459 (2003).
\bibitem{rhewSSE02} J.-H. Rhew, Z. Ren, and M. S. Lundstrom, Solid State
Electronics {\bf 46}, 1899 (2002).
\bibitem{experiments} E. D. Grann {\em et al.}, Phys. Rev. B {\bf 51},
  1631 (1995); E. D. Grann {\em et al.}, Phys. Rev. B {\bf 53},
  9838 (1996); K. T. Tsen {\em et al.}, Appl. Phys. Lett. {\bf 69},
  3575 (1996); W. Liang {\em et al.}, Appl. Phys. Lett. {\bf 82}, 1413
  (2003); W. Liang {\em et al.}, Appl. Phys. Lett. {\bf 84}, 3681
  (2004).  
\bibitem{baranger} H. U. Baranger, and J. W. Wilkins,
  Phys. Rev. B {\bf 36}, 1487 (1987); H. U. Baranger, and J. W. Wilkins,
  Phys. Rev. B {\bf 30}, 7349 (1984).
\bibitem{parameters} GaAs material parameters: $m^{\ast}=0.067m_{0}$,
  $\epsilon=13.1$, $\epsilon_{\infty}=10.92$, $\hbar
  \omega_{0}=0.03536$ eV. 
\bibitem{csontosJCE04} D. Csontos and S. E. Ulloa, (to be published in
J. Comp. Electronics). Also available at arXiv.org, cond-mat/0411499.
\bibitem{scattering} $e^{-}-e^{-}$ scattering, while possibly comparable to 
POP scattering in the $n^{+}$ regions, is much weaker in the channel 
region due to the low doping ($N_{D}=10^{13}$ cm$^{-3}$).
\bibitem{nag} B. R. Nag, {\it Theory of electrical transport in
  semiconductors} (Pergamon Press, Oxford, 1972).
\bibitem{sanoPRL04} N. Sano, Phys. Rev. Lett. {\bf 93}, 246803 (2004).
\bibitem{sanoAPL04} N. Sano, Appl. Phys. Lett. {\bf 85}, 4208 (2004).
\end{thebibliography}
\end{document}